\documentclass[12pt,oneside]{article}
\usepackage{amsfonts}
\usepackage{eufrak}
\usepackage[T1]{fontenc}
\title{On the Khalfin's improvement\\ of the LOY
effective Hamiltonian \\for neutral meson complex}
\author{J. Jankiewicz\footnote{email: jjank@proton.if.uz.zgora.pl}, K.
Urbanowski\footnote{email: K.Urbanowski@proton.if.uz.zgora.pl}\\
\small{University of Zielona G\'ora, Institute of Physics}\\
\small{ul. Prof. Z. Szafrana 4a,  65-516 Zielona G\'ora, Poland}}

\begin{document}
\bibliographystyle{plain}
\maketitle {\noindent}{\em PACS numbers:} 03.65.Ca., 11.10.St.,
11.30.Er., 13.20.Eb., 13.25.Es.,  
\hfill\\
\begin{abstract}
The general properties of the effective Hamiltonian for neutral
meson system improved by L.A. Khalfin in 1980 are studied. It is
shown that contrary to the standard result of the Lee--Oehme--Yang
(LOY) theory, the diagonal matrix elements of this effective
Hamiltonian can not be equal in a CPT invariant system. It is also
shown that the scalar product of short, $|K_{S}\rangle$, and long,
$|K_{L}\rangle$, living superpositions of neutral kaons can not be
real when CPT symmetry is conserved in the system under
considerations whereas within the LOY theory such a scalar product
is real.
\end{abstract}

\section{Introduction}

A formalism convenient for a description of the properties of
neutral $K$ mesons and their time evolution was proposed by Lee,
Oehme and Yang (LOY) in 1956 \cite{j2:LOY}. Within the LOY approach
the Weisskopf--Wigner (WW) approximation used for studying the time
evolution of a single quasistationary state \cite{j2:WW} was adapted
to the case of two state (two particle) subsystems. Within the WW
and LOY approaches it is assumed that time evolution of the total
system under consideration containing one or two quasistationary
states is governed by the Schr\"{o}dinger equation
\begin{equation}
i  \  \frac{\partial | \psi (t) \rangle}{\partial t}= H \ | \psi (t)
\rangle, \ \ \ \ \ \ | \psi (t=0) \rangle =| \psi_{0} \rangle,
\label{j2-Sch}
\end{equation}
where $H$ is the selfadjoint Hamiltonian of the total system and
$|\psi (t) \rangle, |\psi_{0}\rangle $ are vectors belonging to the
total state space $ \mathfrak{H}$ of the system. Using the
interaction representation LOY found approximate solutions of time
dependent equation (\ref{j2-Sch}) for two state problem and then the
Schr\"{o}dinger--like equation with nonhermitian effective
Hamiltonian ${\cal H}_{||}$ governing the the time evolution of
these two states (see \cite{j2:LOY,j2:Gaillard}). Unfortunately not
all steps of  the approximations applied in
\cite{j2:LOY,j2:Gaillard} to obtain the solution of the problem are
well defined.  Some attempts to give  a more exact derivation of the
equation governing time evolution of two state subsystem and to
improve the LOY result were based on the approach exploited in
\cite{j2:Goldberger} for studying properties of unstable states. As
an example of such attempts one can consider the method described in
\cite{j2:Bilenkij,j2:Khalfin}. This method reproduces not only the
LOY effective Hamiltonian ${\cal H}_{||}$ but it also enables one to
relatively simply improve the LOY effective Hamiltonian. It appears
that an examination of the formulae for matrix elements of the
improved ${\cal H}_{||}$ obtained in \cite{j2:Khalfin} suggests that
those formulae contain some inconsistences. The Khalfin's method to
improve the effective Hamiltonians  for multi--component systems is
elegant and seems to be important as it has the potential of
correctly describing the process in question. This is why we decided
to use it to find the exact expressions  for the matrix elements of
Khalfin's effective Hamiltonian. The aim of this paper is to give a
detailed analysis of the approximation described and exploited in
\cite{j2:Bilenkij,j2:Khalfin} and to compare the properties of the
effective Hamiltonians obtained within this approximation with other
effective Hamiltonians. In sec. 2 we review briefly the method used
in \cite{j2:Bilenkij} and in first two Sections of \cite{j2:Khalfin}
to obtain ${\cal H}_{||}$. In Sec. 3 we describe an improved
effective Hamiltonian for neutral $K$ complex derived by Khalfin in
Sec. 3 of his paper \cite{j2:Khalfin} and we give the formulae for
matrix elements of this Hamiltonian free from the above mentioned
inconsistencies. In Sec. 4 we study and discuss some properties of
the mentioned improved effective Hamiltonian not considered in
\cite{j2:Khalfin}. Sec. 5 contains a short discussion of properties
of the Khalfin's improved effective Hamiltonian as well as final
remarks.

\section{Neutral kaons within Weisskopf--Wigner\\ Approximation}

Let us follow  \cite{j2:Bilenkij,j2:Khalfin} and  denote by  $H$ a
self--adjoint Hamiltonian for the total physical system containing a
neutral meson subsystem and assume that
\begin{equation}
 H=H^{o}+H^{w}, \label{j2-1b}
\end{equation}
where $ H^{o}$ denotes the sum of strong and electromagnetic
interactions and $H^{w}$ stands for weak interactions. Next it is
assumed that the operator $H^{o}$ has a complete set of eigenvectors
$\{ |K^{0} \rangle ,|\bar{K}^{0} \rangle, |F \rangle \}$,  where
\begin{equation}
H^{o} \ |K^{0} \rangle = m_{K^{0}} \ |K^{0} \rangle, \ \ \ \ \ H^{o}
\ |\bar{K}^{0}\rangle = m_{\bar{K}^{0}}  \ | \bar{K}^{0} \rangle,
 \ \ \ \ \ H^{o}\ |F \rangle = E_{F}\ |F\rangle,
\label{j2-5b}
\end{equation}
and
\[
\langle K^{0}| \bar{K}^{0} \rangle=0,\;\;\;\; \langle K^{0}| {K}^{0}
\rangle = \langle \bar{K}^{0}| \bar{K}^{0} \rangle =1, \;\;\;
\langle K^{0}|F \rangle = \langle \bar{K}^{0}| F \rangle =0.
\]
Vectors $|K^{0}\rangle,  \bar{K}^{0} \rangle$ are identified with
the state vectors of neutral $K$ and anti--$K$ mesons. Vectors $|F
\rangle$ correspond with decay products of neutral koans.

Usually it is assumed that the strong and electromagnetic
interactions preserve the strangeness $S$, i.e. that
\begin{equation}
[H^{o},S]=0, \label{j2-4}
\end{equation}
and the same assumption is used in \cite{j2:Bilenkij,j2:Khalfin}.
Vectors $|K^{0}\rangle,  \bar{K}^{0} \rangle$ are the eigenvectors
of the operator $S$:
\begin{equation}
S|K^{0}\rangle=(+1) \ |{K}^{0}\rangle, \ \ \ \ \
S|\bar{K}^{0}\rangle=(-1) \ |\bar{K}^{0}\rangle, \label{j2-5a}
\end{equation}

In \cite{j2:Bilenkij,j2:Khalfin} it is also assumed that the strong
and electromagnetic interactions  are CPT--invariant:
\begin{equation}
[H^{o},{\cal{CPT}} ] = 0 ,  \label{j2-2}
\end{equation}
and that they preserve CP symmetry
\begin{equation}
[H^{o},{\cal{CP}} ] = 0.  \label{j2-3}
\end{equation}
Here ${\cal  C},  \; {\cal P}$ and ${\cal T}$ denote operators
realizing charge conjugation, parity and  time  reversal
respectively, for vectors in  $\mathfrak{H}$.

The following phases convention is used in
\cite{j2:Bilenkij,j2:Khalfin}
\begin{equation}
{\cal{CPT}} \ |\bar{K}^{0}\rangle=|K^{0}\rangle. \label{j2-6}
\end{equation}
This last relation and CPT invariance of $H^{o}$ (\ref{j2-2}) imply
that
\begin{equation}
m_{K^{0}}=\langle K^{0} | H^{o} | K^{0}\rangle = m_{\bar{K}^{0}}=
\langle \bar{K}^{0} | H^{o} |\bar{K}^{0} \rangle=m. \label{j2-7}
\end{equation}

Authors of \cite{j2:Bilenkij,j2:Khalfin}  deriving the effective
Hamiltonian governing the time evolution in the subspace of states
spanned by vectors $|K^{0}\rangle,  |\bar{K}^{0} \rangle$ start from
the solution of the Schr\"{o}dinger Equation (\ref{j2-Sch}) for $t
> t_{0} =0$ having following the form
\begin{equation}
|\psi (t) \rangle = - \frac{1}{2 \pi i}  \  \int_{- \infty}^{\infty}
e^{\textstyle -i E t} \ G_{+}(E) \ d E  \ | \psi_{0} \rangle,
\;\;\;\;\;\; (t > 0). \label{j2-9}
\end{equation}
where
\begin{equation}
G_{+}(E)=  (E-H+i \varepsilon )^{-1}. \label{j2-10}
\end{equation}
Using (\ref{j2-1b}) one finds \cite{j2:Goldberger}
\begin{equation}
G_{+}(E)\equiv (E-H^{o}+i \varepsilon)^{-1}+(E-H^{o}+i
\varepsilon)^{-1} \ H^{w} \  G_{+}(E). \label{j2-11}
\end{equation}

Solutions $|\psi (t) \rangle$ of Eq. (\ref{j2-Sch}) can be expanded
into a complete set of eigenvectors $\{|K^{0} \rangle ,|\bar{K}^{0}
\rangle ,|F \rangle \}$ of the operator $H^{o}$.  For the problem
considered it is assumed that there is
\begin{equation}
| \psi_{0} \rangle = a_{K^{0}}(0) \ |K^{0}\rangle +
a_{\bar{K}^{0}}(0) \ |\bar{K}^{0} \rangle  \label{j2-12a}
\end{equation}
at the initial instant $t=t_{0}=0$. So there are no decay products
in the system at $t_{0}=0$. They can be detected there only at the
instant $t>t_{0}=0$. Thus
\begin{equation}
| \psi (t) \rangle = a_{K^{0}}(t) \ |K^{0} \rangle+
a_{{\bar{K}}^{0}}(t) \ |\bar{K}^{0} \rangle + \sum_{F} \sigma_{F}(t)
\ |F \rangle \stackrel{\rm def}{=} |\psi (t)\rangle_{\parallel} +
|\psi (t)\rangle_{\perp} \label{j2-12b}
\end{equation}
at $t > t_{0} =0$ and therefore $ \sigma_{F}(0)=0$ and $
\sigma_{F}(t) \neq 0$ for $t
> 0$. Here
\[
|\psi (t)\rangle_{\parallel} = a_{K^{0}}(t) \ |K^{0} \rangle+
a_{{\bar{K}}^{0}}(t) \ |\bar{K}^{0} \rangle \;\;\;\;\; {\rm
and}\;\;\;\;\; |\psi (t)\rangle_{\perp} = \sum_{F} \sigma_{F}(t) \
|F \rangle  \label{psi||}
\]
and $|\psi (t)\rangle_{\parallel} \in \mathfrak{H}_{||} \subset
\mathfrak{H}$, $|\psi (t)\rangle_{\perp} \in \mathfrak{H} \ominus
\mathfrak{H}_{||}$. The subspace $\mathfrak{H}_{||}$ is a two
dimensional subspace of state space $\mathfrak{H}$ spanned by
orthogonal vectors $|K_{0} \rangle, |\bar{K}_{0}\rangle$.

We have
\begin{equation}
 a_{K^{0}}(t)= \langle K^{0}| \psi(t)\rangle \equiv
 \langle K^{0}| \psi(t)\rangle_{\parallel} , \;\;
 a_{\bar{K}^{0}}(t)= \langle \bar{K}^{0}| \psi(t)\rangle
 \equiv \langle \bar{K}^{0}| \psi(t)\rangle_{\parallel}. \label{j2-13}
\end{equation}
From (\ref{j2-9}) one infers that
\begin{equation}
a_{\alpha}(t) = - \frac{1}{2 \pi i}  \  \int_{- \infty}^{\infty}
e^{\textstyle -i E t} \ \sum_{\beta} \langle \alpha | \ G_{+}(E) \ |
\beta \rangle \ a_{\beta}(0) \ dE, \;\;\;\;\;\; (t > 0),
\label{j2-a(t)}
\end{equation}
where $\alpha, \beta = K^{0}, \bar{K}^{0}$. Using relation
(\ref{j2-11}) and  condition (\ref{j2-7}) after some algebra one can
rewrite  solutions (\ref{j2-a(t)}) of Eq. (\ref{j2-Sch}) in the
matrix form as follows
\begin{equation}
a(t) = -\frac{1}{2 \pi i} \  \int_{- \infty}^{\infty} e^{\textstyle
-i E t} \ \frac{d E}{E-m-R(E)+i \varepsilon} \  a(0), \label{j2-14}
\end{equation}
where $(t > 0)$ and $a(t)$ i $a(0)$ are one column matrices
\begin{equation} a(t)= \left( \begin{array}{cc}
a_{K^{0}}(t) \\ a_{\bar{K}^{0}}(t)
\end{array} \right), \;\;\;\;\;
a(0)= \left( \begin{array}{cc} a_{K^{0}}(0) \\ a_{\bar{K}^{0}}(0)
\end{array} \right), \label{j2-15}
\end{equation}
and $R(E)$ is  $(2 \times 2)$ matrix with matrix elements $R_{\alpha
\beta}(E)$, where $\alpha,\beta = K^{0},\bar{K}^{0}$ and
\begin{eqnarray}
\nonumber R_{\alpha \beta}(E)&=& \langle \alpha|H^{w}| \beta \rangle
\nonumber
\\&&+ \sum_{F} \langle \alpha|H^{w}|F \rangle  \ \frac{1}{E-E_{F}+i
\varepsilon} \ \langle
F|H^{w}|\beta \rangle  \nonumber \\
&& +\sum_{F} \sum_{F'} \Big\{ \langle \alpha|H^{w}|F \rangle
 \ \frac{1}{E-E_{F}+i \varepsilon} \ \times \nonumber \\
 && \times \langle F|H^{w}|F' \rangle
 \ \frac{1}{E-E_{F'}+i \varepsilon} \
  \langle F'|H^{w}|\beta \rangle  \Big\} \ +...
\label{j2-R-ab} \\
&& \equiv \langle \alpha|H^{w}|\beta \rangle - \Sigma_{\alpha
\beta}(E), \label{j2-18}
\end{eqnarray}
and
\begin{eqnarray}
\Sigma_{\alpha \beta}(E)&=& \sum_{F, F'} \langle \alpha|H^{w}|F
\rangle \langle F | \ \frac{1}{QHQ-E-i 0} \ |F'\rangle
 \langle F'|H^{w}|\beta \rangle \nonumber \\
 &\equiv & \langle \alpha|H^{w} \ Q \ \frac{1}{QHQ-E-i 0} \ Q \
 H^{w}|\beta
 \rangle,  \label{j2-18a}\\
 & \equiv & \langle \alpha| \ \Sigma (E) \ |\beta
 \rangle, \label{j2-18b}
\end{eqnarray}
and
\begin{eqnarray}
\Sigma (E) &=& PHQ \ \frac{1}{QHQ-E-i 0}  \ QHP, \label{j2-20b1}\\
 Q&=&\sum_{F} |F \rangle \langle F|, \label{j2-20b}\\
 P & = & \mathbb{I} - Q \equiv |K^{0} \rangle\langle K^{0}| \ +
  \ |\bar{K}^{0}
 \rangle\langle \bar{K}^{0}|. \label{j2-20b2}
\end{eqnarray}
The relation (\ref{j2-14}) is the exact formula for solutions of Eq.
(\ref{j2-Sch}) for $t > 0$. The problem is how to evaluate the
integral (\ref{j2-14}) and thus the amplitudes $a_{\alpha}(t)$.
Usually it is possible within the use of some approximate methods.
Depending on the methods used one obtains more or less accurate
expressions for $a_{\alpha}(t)$ and therefore more or less accurate
description of the properties of the physical system considered.

From experimental data it is known that
\begin{eqnarray}
\langle K^{0}|H^{o}|K^{0}\rangle = m  \ \gg \ \langle
K^{0}|H^{w}|K^{0}\rangle = \Delta_{w}m. \label{j2-22}
\end{eqnarray}
This enables one to assume that $|R_{\alpha \beta}(E)| \ \ll m$. So
the conclusion that the position of the pole of the expression under
the  integral in (\ref{j2-14}) is very close to $m$, seems to be
reasonable. Thus one can expect that  replacing $R(E)$ by $R(m)$ in
(\ref{j2-14}) should not cause a large deviation from the exact
value of this integral. Making use of this conclusion the value of
the integral (\ref{j2-14}) can be computed within so--called
Weisskopf--Wigner
approximation \cite{j2:WW,j2:Bilenkij,j2:Khalfin}.\\

According to \cite{j2:Khalfin} the WW approximation consists in:
\begin{enumerate}
    \item Taking into account only the pole contribution into the
    value of the integral (\ref{j2-14}) (i.e., a neglecting all the
    cut  and threshold, etc., contributions into the value of the integral
    (\ref{j2-14})).
    \item Replacing $R(E)$ by its value for $E = m$, (ie. inserting $R(m)$
instead of $R(E)$ in (\ref{j2-14}) ).
\end{enumerate}
(Note that it is rather difficult to find an exact estimation of the
error generated by such a procedure). Applying this prescription to
the integral (\ref{j2-14}) yields
\begin{eqnarray}
a(t) &\simeq& a^{WW}(t) =  -\frac{1}{2 \pi i} \ \int_{-
\infty}^{\infty} e^{\textstyle -i E t} \ \frac{d E}{E-m-R(m)+i
\varepsilon} \  a(0) \label{j2-a-WW} \\
&\equiv& e^{\textstyle -i \mathcal{H}^{WW}t} \ a(0), \label{j2-23}
\end{eqnarray}
for $(t>0)$, where
\begin{equation}
\mathcal{H}^{WW} \stackrel{\rm def}{=} m \
\mathbb{I}_{\parallel}+R(m)\equiv M^{WW} - \frac{i}{2} \ \Gamma^{WW}
\label{j2-H-ww}
\end{equation}
is an nonhermitian operator, $M^{WW} = (M^{WW})^{+}, \ \Gamma^{WW} =
(\Gamma^{WW})^{+}$ and
\begin{equation}
\mathbb{I}_{\parallel} = \left(%
\begin{array}{ccc}
  1 &,& 0 \\
  0 &,& 1 \\
\end{array}%
\right). \label{j2-23a}
\end{equation}
Matrix elements of $\mathcal{H}^{WW}$ have the following form
\begin{eqnarray}
\mathcal{H}^{WW}_{\alpha \beta} &=& m \ \delta_{\alpha \beta} +
R_{\alpha \beta} (m), \label{j2-H-WW-ab1}\\
&\equiv& m \ \delta_{\alpha \beta} \  +  \ H^{w}_{\alpha \beta} -
\Sigma_{\alpha \beta} (m), \label{j2-H-WW-ab2}
\end{eqnarray}
where $ \alpha ,\beta = K^{0}, \bar{K}^{0}$ and $H_{\alpha
\beta}^{w} = \langle \alpha|H^{w}|\beta \rangle$.

All operators appearing in the definition (\ref{j2-H-ww}) act in two
dimensional subspace of states $\mathfrak{H}_{||}$. From
(\ref{j2-23}) it follows that $a(t) \simeq a^{WW}(t) \in
\mathfrak{H}_{||}$ solves the following Schr\"{o}dinger--like
equation
\begin{equation}
i  \ \frac{\partial a^{WW}(t)}{\partial t} = \mathcal{H}^{WW} \ \
a^{WW}(t) \label{j2-25}
\end{equation}
where the operator $\mathcal{H}^{WW}$ is the effective Hamiltonian
for vectors belonging to the subspace $\mathfrak{H}_{||}$.  Matrix
elements of $\mathcal{H}^{WW}$ are defined by formulae
(\ref{j2-18}). They are exactly the same as those obtained by LOY
(see \cite{j2:LOY,j2:Gaillard,j2:5,j2:6}).

If assumption (\ref{j2-2}) is completed by the following one
\begin{equation}
[H,CPT]=0,  \label{j2-CPT-H}
\end{equation}
that is, if one assumes that the system containing neutral kaons is
CPT invariant, then using relations (\ref{j2-6}), (\ref{j2-7}) one
easily finds from (\ref{j2-18}) that
\begin{equation}
R_{K^{0} K^{0}}(m)=R_{\bar{K}^{0} \bar{K}^{0}}(m) \label{j2-R11=22}
\end{equation}
and thus
\begin{equation}
\mathcal{H}^{WW}_{K^{0} K^{0}}=\mathcal{H}^{WW}_{\bar{K}^{0}
\bar{K}^{0}}, \, \, \, M^{WW}_{K^{0} K^{0}}=M^{WW}_{\bar{K}^{0}
\bar{K}^{0}}, \, \, \, \Gamma^{WW}_{K^{0} K^{0}}=
\Gamma^{WW}_{\bar{K}^{0} \bar{K}^{0}}. \label{j2-26}
\end{equation}
These relations are the standard conclusions of the LOY theory for
CPT invariant physical systems \cite{j2:Gaillard,j2:Bilenkij},
\cite{j2:5}
--- \cite{j2:branco}.

Assuming that interactions $H^{w}$ responsible for the decay
processes in the system considered violate CP symmetry,
$[{\cal{CP}}, H] \neq 0$ one finds another important result of the
LOY approach;
\begin{equation}
R_{K^{0} \bar{K}^{0}}(m)\neq R_{\bar{K}^{0} K^{0}}(m),
\label{CP-R12}
\end{equation}
which implies that
\begin{equation}
\mathcal{H}^{WW}_{K^{0} \bar{K}^{0}} \neq
\mathcal{H}^{WW}_{\bar{K}^{0} K^{0}}, \, \, \, M^{WW}_{K^{0}
\bar{K}^{0}} \neq M^{WW}_{\bar{K}^{0} K^{0}}, \, \, \,
\Gamma^{WW}_{K^{0} \bar{K}^{0}} \neq \Gamma^{WW}_{\bar{K}^{0}
K^{0}}. \label{j2-28}
\end{equation}

The eigenvalue equations for $\mathcal{H}^{WW}$ have the following
form \cite{j2:Bilenkij,j2:Khalfin}
\begin{equation}
\mathcal{H}^{WW} \ a^{WW}_{S}= \lambda_{S} \ a^{WW}_{S}, \ \ \ \ \ \
\ \mathcal{H}^{WW} \ a^{WW}_{L}= \lambda_{L} \ a^{WW}_{L},
\label{j2-29}
\end{equation}
where  $a^{WW}_{S}$ and $a^{WW}_{L}$ are eigenfunctions of the
operator $\mathcal{H}^{WW}$ for the eigenvalues
\begin{equation}
\lambda_{S} = m_{S} - \frac{i}{2} \ \Gamma_{S}, \;\;\;\; \lambda_{L}
= m_{L} - \frac{i}{2} \ \Gamma_{L}. \label{j2-LS-LL}
\end{equation}

In the case of CPT invariant system, (\ref{j2-CPT-H}), relations
(\ref{j2-26}) hold which leads to the following form of eigenvectors
for $\mathcal{H}^{WW}$ \cite{j2:Khalfin},
\begin{eqnarray}
|K_{S}\rangle \equiv \frac{1}{p  \
(1+|\frac{q}{p}|^2)^{\frac{1}{2}}}  \  (p \  |K^{0}\rangle  \ - q \
|\bar{K}^{0}\rangle) \label{j2-32}
\end{eqnarray}
and
\begin{eqnarray}
|K_{L}\rangle \equiv \frac{1}{p  \
(1+|\frac{q}{p}|^2)^{\frac{1}{2}}}  \  (p \  |K^{0}\rangle \ + q \
|\bar{K}^{0}\rangle), \label{j2-33}
\end{eqnarray}
or, equivalently,
\begin{eqnarray}
|K_{S}\rangle = \rho^{WW}\bigg(|K^{0} \rangle-r |\bar{K}^{0}
\rangle\bigg), \label{j2-32a}
\end{eqnarray}
\begin{eqnarray}
|K_{L}\rangle = \rho^{WW}\bigg(|K^{0} \rangle+r |\bar{K}^{0}
\rangle\bigg), \label{j2-33a}
\end{eqnarray}
where
\begin{eqnarray}
q \equiv \sqrt{ \mathcal{H}^{WW}_{\bar{K}^{0} K^{0}}} \ , \ \ \ p
\equiv \sqrt{ \mathcal{H}^{WW}_{K^{0} \bar{K}^{0}}}.  \label{j2-31}
\end{eqnarray}
and
\begin{eqnarray}
r= \frac{q}{p} = \sqrt{ \frac{ {\mathcal{H}_{ \bar{K}^{0}
K^{0}}^{WW}}} { {\mathcal{H}_{K^{0} \bar{K}^{0}} ^{WW}}}}.
\label{j2-33b}
\end{eqnarray}
So within the WW approximation physical states of neutral kaons,
$K_{S}$ and $K_{L}$, are linear superpositions of $K^{0}$ and
$\bar{K}^{0}$ and they decay exponentially evolving in time in
$\mathfrak{H}_{||}$ (see (\ref{j2-23}) and (\ref{j2-29})),
\begin{eqnarray} |K_{S} (t) \rangle_{\parallel} = e^{\textstyle - i
\mathcal{H}^{WW} t} \ |K_{S}\rangle = e^{\textstyle -i(m_{S}-
\frac{i}{2} \ \Gamma_{S})t} \ |K_{S}\rangle \in  \mathfrak{H}_{||},
\label{j2-34}
\end{eqnarray}
\begin{eqnarray}
|K_{L} (t) \rangle_{\parallel} = e^{\textstyle -i \mathcal{H}^{WW}
t} \ |K_{L}\rangle = e^{\textstyle -(im_{L} -\frac{i}{2} \
\Gamma_{L})t} \ |K_{L}\rangle \in  \mathfrak{H}_{||}, \label{j2-34}
\label{j2-35}
\end{eqnarray}

These last two relations and (\ref{j2-32a}), (\ref{j2-33a}) enable
one to determine the time evolution  of vectors $|K^{0}\rangle$ and
$|\bar{K}^{0} \rangle$ in $\mathfrak{H}_{||}$. One finds, eg., that
within the WW approximation
\begin{eqnarray}
|K^{0}(t)\rangle_{\parallel} &\simeq& |K^{0}_{WW}(t)
\rangle_{\parallel} \nonumber =
e^{\textstyle - i \mathcal{H}^{WW} t}  \;  |K^{0}\rangle  \nonumber \\
&=&  \;\;\; \;\frac{1}{2} \  \biggl[e^{\textstyle -i(m_{L}-
\frac{i}{2} \ \Gamma_{L})t} \ + \ e^{\textstyle -i(m_{S}-
\frac{i}{2} \ \Gamma_{S})t} \biggl] \  |K^{0} \rangle  \nonumber \\
&&\;+\frac{r}{2} \  \biggl[e^{\textstyle -i(m_{L} - \frac{i}{2} \
\Gamma_{L})t} \ - \ e^{\textstyle -i(m_{S}- \frac{i}{2} \
\Gamma_{S})t} \biggl] \ |\bar{K}^{0} \rangle . \label{j2-36}
\end{eqnarray}

From the relations (\ref{j2-32}) and (\ref{j2-33}) follows another
fundamental conclusion of the LOY theory about properties CPT
invariant systems of neutral mesons. Namely physical states $K_{L},
K_{S}$ have the form of (\ref{j2-32}) and (\ref{j2-33}) only if the
condition (\ref{j2-CPT-H}) and thus properties (\ref{j2-26}) hold.
It easy to calculate the scalar product of the vectors
$|K_{S}\rangle, |K_{L}\rangle$ and one finds that
\begin{equation}
\langle K_{S}|K_{L} \rangle = \frac{|p|^{2} - |q|^{2}}{|p|^{2} +
|q|^{2}} = (\langle K_{S}|K_{L} \rangle)^{\ast} = \langle
K_{L}|K_{S} \rangle \, \neq \,0 . \label{j2-KS-KL}
\end{equation}
This property means that within the LOY theory (i.e., within the WW
approximation) the imaginary part, $\Im \,(\langle K_{S}|K_{L}
\rangle)$, of the product $\langle K_{S}|K_{L} \rangle$ can be
considered as the measure of a possible violation of the CPT
symmetry: LOY theory states that system considered is CPT invariant
only if $\Im \,(\langle K_{S}|K_{L} \rangle) \, =\,0$ (see:
\cite{j2:Gaillard,j2:Bilenkij}, \cite{j2:5}
--- \cite{j2:branco}).

\section{Khalfin's effective Hamiltonian for neutral kaon complex}

In \cite{j2:Khalfin} an observation is made that the approximation
$R(E) \simeq R(m)$ is not the best and leads to an indeterminate
error in evaluating the amplitude $a(t)$, (\ref{j2-14}). It is
obvious that using the more accurate estimation of $R(E)$ should
yield a more accurate formula for $a(t)$. The suggestion is made in
\cite{j2:Khalfin} that the more accurate approximation for $R(E)$
can be obtained expanding $R(E)$ into its Taylor series expansion
around the point $E = m$. This idea gives
\begin{eqnarray}
R(E) = R(m)+(E-m) \  \frac{d R(E)}{d E} \Biggl{|}_{E=m} \ + \
\frac{(E - m)^{2}}{2} \ \frac{d^{2} R(E)}{d E^{2}} \Biggl{|}_{E=m} \
+ \ldots , \label{j2-Taylor}
\end{eqnarray}
So, the minimal improvement of the approximation $R(E) \simeq R(m)$
is the following one \cite{j2:Khalfin}
\begin{eqnarray}
R(E)\simeq R(m)+(E-m) \  \frac{d R(m)}{d m}, \label{j2-40}
\end{eqnarray}
where
\begin{eqnarray}
\frac{d R(m)}{d m} \ \equiv \  \frac{d R(E)}{d E}\Biggl{|}_{E=m}.
\label{j2-41}
\end{eqnarray}
Having this improved estimation of $R(E)$ and looking for more
accurate expression for the amplitude $a(t)$ the WW approximation
defined in the previous Section is modified in \cite{j2:Khalfin} by
ignoring the point 2 in the mentioned definition and replacing the
approximation $R(E) \simeq R(m)$ used there by the relation
(\ref{j2-40}). This procedure is the Khalfin's improvement of the WW
approximation.

Using (\ref{j2-40}) the denominator of the expression  under the
integral in  formula (\ref{j2-14}) for the amplitude $a(t)$ takes
the following form
\begin{eqnarray}
E-m-R(E)\simeq \Biggl(1-\frac{d R(m)}{d m}\Biggl)  \
\Biggl[E-m-\Biggl(1-\frac{d R(m)}{d m}\Biggl)^{-1}  \ R(m)\Biggl].
\label{j2-42}
\end{eqnarray}
Inserting (\ref{j2-42}) into (\ref{j2-14})  yields
\begin{eqnarray}
 a(t) & \simeq & \tilde{a}(t) = -\frac{1}{2 \pi i}  \  \int_{-
\infty}^{\infty} dE \Big\{e^{\textstyle -i E t} \times \nonumber \\
&&\;\;\;\;\;\;\;\;\;\;\times \frac{1}{E-m-\biggl(1-\frac{d R(m)}{d
m}\biggl)^{-1} \
R(m)+i \varepsilon} \ \Big\} \ \times \nonumber \\
&&\;\;\;\;\;\;\;\;\;\;\times \Biggl(1-\frac{d R(m)}{d m}\Biggl)^{-1}
\ a(0), \;\;\;\;\;\;\; (t >0), \label{j2-43}
\end{eqnarray}
This expression for $a(t)$ replaces and improves  the approximate
formula (\ref{j2-a-WW}) for $a(t) \simeq a^{WW}(t)$.

Taking into account only the pole contribution into the value of the
integral (\ref{j2-43}) leads to the result (see \cite{j2:Khalfin})
\begin{eqnarray}
a(t) \simeq  \tilde{a}(t) = e^{\textstyle -i \tilde{\mathcal{H}} t}
  \  \tilde{a}(0), \;\;\;\;(t >0),\label{j2-44}
\end{eqnarray}
where
\begin{equation}
\tilde{a}(0) \stackrel{\rm def}{=} \mathbf{A} \ a(0) \ \equiv \
\biggl(1-\frac{d R(m)}{d m}\biggl)^{-1}  \  a(0), \label{a-0-tilde}
\end{equation}
and
\begin{eqnarray}
\tilde{\mathcal{H}} \equiv  m \ \mathbb{I}_{\parallel}+
\biggl(1-\frac{d R(m)}{d m}\biggl)^{-1}  R(m) = \tilde{M} -
\frac{i}{2} \ \tilde{\Gamma}, \label{j2-47}
\end{eqnarray}
(where $\tilde{M} = {\tilde{M}}^{+}$ and $\tilde{\Gamma} =
{\tilde{\Gamma}}^{+}$) denotes the Khalfin's improved effective
Hamiltonian acting in the subspace $\mathfrak{H}_{||}$.

Note that $\tilde{a}(t)$ solves the following, similar to
(\ref{j2-25}),
 Schr\"{o}dinger--like equation
\begin{equation}
i \ \frac{\partial \tilde{a}(t)}{\partial t} \ = \
\tilde{\mathcal{H}} \ \tilde{a}(t). \label{LOY-improved-eq}
\end{equation}
This is the evolution equation for the the subspace of states
$\mathfrak{H}_{||}$ of neutral mesons. One should expect that
solutions $\tilde{a}(t)$ of this equation with the improved
$\tilde{\mathcal{H}}$ will lead to a more accurate description of
real properties of neutral mesons than the solutions $a^{WW}(t)$,
(\ref{j2-23}), of the equation (\ref{j2-25}).

We have
\begin{eqnarray}
\mathbf{A} =  \biggl(1 - \frac{d R(m)}{d m}\biggl)^{-1} = D \
\left(
\begin{array}{ccc}
1-\frac{d R_{\bar{K}^{0} \bar{K}^{0}}(m)}{d m} & , &
\frac{d R_{K^{0} \bar{K}^{0}}(m)}{d m} \\
\frac{d R_{\bar{K}^{0} K^{0}}(m)}{d m} & , &  1-\frac{d R_{K^{0}
K^{0}}(m)}{d m}
\end{array} \right), \label{j2-45}
\end{eqnarray}
where
\begin{eqnarray}
D \nonumber &=& \nonumber \biggl[\det \biggl(1-\frac{d R(m)}{d
m}\biggl)\biggl]^{-1} \nonumber \\
&\equiv& \biggl[1-\frac{d R_{K^{0} K^{0}}(m)}{d m} -\frac{d
R_{\bar{K}^{0} \bar{K}^{0}}(m)}{d m} + \frac{d R_{K^{0}
K^{0}}(m)}{d m}  \  \frac{d R_{\bar{K}^{0} \bar{K}^{0}}(m)}{d m}  \nonumber \\
&& - \frac{d R_{K^{0} \bar{K}^{0}}(m)}{d m}  \  \frac{d
R_{\bar{K}^{0} K^{0}}(m)}{d m} \biggl]^{-1} . \label{j2-46}
\end{eqnarray}
Taking into account (\ref{j2-45}) one infers from (\ref{j2-47}) that
the matrix elements of Khalfin's effective Hamiltonian,
$\tilde{\mathcal{H}}$, have the following form
\begin{eqnarray}
\tilde{\mathcal{H}}_{\alpha \alpha}&=& m+D  \  \biggl(R_{\alpha
\alpha}(m)-R_{\alpha \alpha}(m) \ \frac{d R_{\beta \beta}(m)}{d m}
\nonumber
\\&& \makebox[85pt]{}+ R_{\beta \alpha}(m) \ \frac{d R_{\alpha
\beta}(m)}{d m}\biggl) \label{j2-48}\\
&\equiv & \tilde{M}_{\alpha \alpha} - \frac{i}{2} \
\tilde{\Gamma}_{\alpha \alpha}, \nonumber
\end{eqnarray}
\begin{eqnarray}
\tilde{\mathcal{H}}_{\alpha \beta} &=& D  \ \biggl(R_{\alpha
\beta}(m)-R_{\alpha \beta}(m)  \ \frac{d R_{\beta \beta}(m)}{d
m}\nonumber \\&& \makebox[60pt]{} + R_{\beta \beta}(m)  \ \frac{d
R_{\alpha
\beta}(m)}{d m}\biggl) \label{j2-51}\\
&\equiv & \tilde{M}_{\alpha \beta} - \frac{i}{2} \
\tilde{\Gamma}_{\alpha \beta}, \nonumber
\end{eqnarray}
where $ \alpha \neq \beta$ and $\alpha, \beta = K^{0}, \bar{K}^{0}$.
These two last formulae differ from those obtained in
\cite{j2:Khalfin} for $\tilde{\mathcal{H}}_{\alpha \beta}$ and
$\tilde{\mathcal{H}}_{\alpha \alpha}$. Strictly speaking the last
components in (\ref{j2-48}), (\ref{j2-51}) and the components
corresponding to them in the Khalfin's formulae for matrix elements
of $\tilde{\mathcal{H}}$ are different. What is more the examination
of the expressions for $\tilde{\mathcal{H}}_{\alpha \alpha},
\tilde{\mathcal{H}}_{\alpha \beta}$ given and discussed in
\cite{j2:Khalfin} shows that the mentioned different components in
Khalfin's formulae for $\tilde{\mathcal{H}}_{\alpha \alpha},
\tilde{\mathcal{H}}_{\alpha \beta}$ are wrong in the general case.
For this reason the formulae for $\tilde{\mathcal{H}}_{\alpha
\alpha}, \tilde{\mathcal{H}}_{\alpha \beta}$ used in
\cite{j2:Khalfin} are incorrect. This means that one can not be sure
that all conclusions drawn in \cite{j2:Khalfin} and following from
the analysis of the properties of $\tilde{\mathcal{H}}_{\alpha
\alpha}, \tilde{\mathcal{H}}_{\alpha \beta}$ obtained there reflect
real properties of neutral meson complexes.

Making use of the expansion $(1 - x)^{-1} = 1+x+x^{2}+x^{3}+...$ for
$|x| < 1$, then taking
\begin{eqnarray}
x \nonumber &=& \nonumber \frac{d R_{K^{0} K^{0}}(m)}{d m} +\frac{d
R_{\bar{K}^{0} \bar{K}^{0}}(m)}{d m} -
\frac{d R_{K^{0} K^{0}}(m)}{d m}  \
\frac{d R_{\bar{K}^{0} \bar{K}^{0}}(m)}{d m} \nonumber \\
&& + \frac{d R_{K^{0} \bar{K}^{0}}(m)}{d m}  \  \frac{d
R_{\bar{K}^{0} K^{0}}(m)}{d m} . \label{j2-46b}
\end{eqnarray}
and assuming that this $x$ fulfils the condition $|x| \ll 1$, the
expression (\ref{j2-46}) for $D$ can be approximated by the
following formula
\begin{eqnarray}
D & \simeq &  1+x \label{j2-46d}\\
&=& \nonumber 1+\frac{d R_{K^{0} K^{0}}(m)}{d m} +\frac{d
R_{\bar{K}^{0} \bar{K}^{0}}(m)}{d m} -
\frac{d R_{K^{0} K^{0}}(m)}{d m}  \
 \frac{d R_{\bar{K}^{0} \bar{K}^{0}}(m)}{d m} \nonumber \\
&& + \frac{d R_{K^{0} \bar{K}^{0}}(m)}{d m}  \  \frac{d
R_{\bar{K}^{0} K^{0}}(m)}{d m} . \label{j2-46c}
\end{eqnarray}

Inserting (\ref{j2-46c}) into (\ref{j2-48}), (\ref{j2-51}) and
keeping in these formulae only expressions of order up to that of
type $R_{\alpha \beta}(m)R_{\alpha' \beta'}(m)$, $R_{\alpha \beta}
\frac{R_{\alpha' \beta'}(m)}{dm}$ (where $\alpha,
\alpha',\beta,\beta' = K^{0},\bar{K}^{0}$) one obtains
\begin{eqnarray}
\tilde{\mathcal{H}}_{\alpha \alpha} \simeq m&+& R_{\alpha
\alpha}(m)+R_{\beta \alpha}(m) \frac{d R_{\alpha \beta}(m)}{d
m}\nonumber \\ &+& R_{\alpha \alpha}(m) \frac{d R_{\alpha
\alpha}(m)}{d m}, \label{j2-48a}
\end{eqnarray}
\begin{eqnarray}
\tilde{\mathcal{H}}_{\alpha \beta} &\simeq & R_{\alpha
\beta}(m)+R_{\beta \beta}(m) \frac{d R_{\alpha \beta}(m)}{d m}
\nonumber \\&&+ R_{\alpha \beta}(m) \frac{d R_{\alpha \alpha}(m)}{d
m}, \label{j2-51a}
\end{eqnarray}
where $ \alpha \neq \beta$.

Some general properties of the matrix elements
$\tilde{\mathcal{H}}_{\alpha \beta}$, (see (\ref{j2-48}),
(\ref{j2-51}) and (\ref{j2-48a}), (\ref{j2-51a}) ),  of
$\tilde{\mathcal{H}}$ follow from symmetry properties of the total
Hamiltonian $H$. Assuming CPT invariance of the system containing
neutral mesons, (\ref{j2-2}), one obtains following relations
\begin{eqnarray}
R_{K^{0} K^{0}}(m)=R_{\bar{K}^{0} \bar{K}^{0}}(m), \ \ \ \ \ \
\frac{d R_{K^{0} K^{0}}(m)}{dm}= \frac{d R_{\bar{K}^{0}
\bar{K}^{0}}(m)}{dm}, \label{j2-52}
\end{eqnarray}
which are analogous to (\ref{j2-R11=22}).

If the system is CP invariant beside relations (\ref{j2-52})  the
following additional ones hold too,
\begin{eqnarray}
R_{K^{0} \bar{K}^{0}}(m)=R_{\bar{K}^{0} K^{0}}(m), \ \ \ \ \ \
\frac{d R_{K^{0} \bar{K}^{0}}(m)}{dm}= \frac{d R_{\bar{K}^{0}
K^{0}}(m)}{dm}. \label{j2-53}
\end{eqnarray}
On the other hand if CP symmetry is violated then relations
(\ref{j2-53}) are not valid and one has
\begin{eqnarray}
R_{K^{0} \bar{K}^{0}}(m) \neq R_{\bar{K}^{0} K^{0}}(m), \ \ \ \ \ \
\frac{d R_{K^{0} \bar{K}^{0}}(m)}{dm} \neq \frac{d R_{\bar{K}^{0}
K^{0}}(m)}{dm}. \label{j2-54}
\end{eqnarray}

Finally, if the system is CPT invariant and CP symmetry is not
conserved there then relations (\ref{j2-52}) and (\ref{j2-54})
occur. This means, by (\ref{j2-48}), that
$\tilde{\mathcal{H}}_{K^{0} K^{0}} \neq
\tilde{\mathcal{H}}_{\bar{K}^{0} \bar{K}^{0}}$ , that is that
\begin{eqnarray}
\tilde{\mathcal{H}}_{K^{0} K^{0}} - \tilde{\mathcal{H}}_{\bar{K}^{0}
\bar{K}^{0}} \neq 0. \label{j2-56}
\end{eqnarray}
So it appears that the minimal improvement of the standard WW
approximation leads to the conclusion that one of the fundamental
results of the LOY theory for CPT invariant systems, ie., the
relation (\ref{j2-26}), can not be considered as universally
valid for real systems.

A conclusion analogous to (\ref{j2-56}) was also obtained in
\cite{j2:Khalfin} but for the reasons mentioned after (\ref{j2-48}),
(\ref{j2-51}) the formula for $(\tilde{\mathcal{H}}_{K^{0} K^{0}} -
\tilde{\mathcal{H}}_{\bar{K}^{0} \bar{K}^{0}})$ obtained therein
differs from that used in the next Section and following from
(\ref{j2-48}), (\ref{j2-51}).

\section{Some properties of Khalfin's improved \\effective Hamiltonian
$\tilde{\mathcal{H}}$}

In this Section we will
assume that CPT symmetry holds in the system under considerations.
In such a case relations (\ref{j2-48}) and (\ref{j2-52}) ---
(\ref{j2-54}) lead to the following expression for the difference of
diagonal matrix elements (\ref{j2-56}) of the Khalfin's effective
Hamiltonian for neutral mesons complex
\begin{eqnarray}
\tilde{\mathcal{H}}_{K^{0} K^{0}} - \tilde{\mathcal{H}}_{\bar{K}^{0}
\bar{K}^{0}}\stackrel{def} {=} 2\tilde{h}_{z} &=& D \biggl(R_{K^{0}
\bar{K}^{0}}(m) \frac{d R_{\bar{K}^{0} K^{0}}(m)}{d m}\nonumber \\
&&\;\;\; - R_{\bar{K}^{0} K^{0}}(m) \frac{d R_{K^{0}
\bar{K}^{0}}(m)}{d m}\biggl) \neq 0. \label{j2-57}
\end{eqnarray}

Sometimes it is convenient to express this difference in terms of
matrix elements $\Sigma_{\alpha \beta}(m)$ instead of $R_{\alpha
\beta}(m)$. Taking into account formula (\ref{j2-18}) for $R_{\alpha
\beta}(m)$ and making use of the fact that
\begin{equation}
\frac{d H^{w}_{\alpha \beta}}{d m}\equiv 0, \label{j2-dHw-dm}
\end{equation}
one finds
\begin{eqnarray}
\frac{d R_{\alpha \beta}(m)}{d m} = - \frac{ d\, \Sigma_{\alpha
\beta}(m)}{d m}. \label{j2-61}
\end{eqnarray}
Inserting (\ref{j2-18}) and (\ref{j2-61}) into (\ref{j2-57}) one
obtains
\begin{eqnarray}
\tilde{\mathcal{H}}_{K^{0} K^{0}} -
\tilde{\mathcal{H}}_{\bar{K}^{0} \bar{K}^{0}} \nonumber &=& D
 \  \biggl(- H^{w}_{K^{0} \bar{K}^{0}}  \
 \frac{d\, \Sigma_{\bar{K}^{0} K^{0}}(m)}{d m}\\
 && + \ H^{w}_{\bar{K}^{0} K^{0}}
 \  \frac{d\, \Sigma_{K^{0} \bar{K}^{0}}(m)}{d m}
 + \Sigma_{K^{0} \bar{K}^{0}}(m)  \  \frac{d\, \Sigma_{\bar{K}^{0}
K^{0}}(m)}{d m} \nonumber \\ && - \Sigma_{\bar{K}^{0} K^{0}}(m)
 \  \frac{d\, \Sigma_{K^{0} \bar{K}^{0}}(m)}{d m} \biggl), \label{j2-63}
\end{eqnarray}
where for $D$, (\ref{j2-46}), due to the property (\ref{j2-R11=22}),
one has
\begin{eqnarray}
D \nonumber &=& \biggl[1 + 2  \  \frac{d\,\Sigma_{K^{0} K^{0}}(m)}{d
m}+
\biggl(- \frac{d\, \Sigma_{K^{0} K^{0}}(m)}{d m}\biggl)^{2}  \nonumber \\
&& - \frac{d\, \Sigma_{K^{0} \bar{K}^{0}}(m)}{d m}  \  \frac{d\,
\Sigma_{\bar{K}^{0} K^{0}}(m)}{d m}\biggl]^{-1}. \label{j2-66}
\end{eqnarray}

Ignoring in (\ref{j2-63}) terms of the form $ \Biggl(\Sigma_{\alpha
\beta}(m)  \  \frac{d \,\Sigma_{\beta \alpha}(m)}{d m}\Biggl),$
where $\alpha \neq \beta$ yields
\begin{eqnarray}
\tilde{\mathcal{H}}_{K^{0} K^{0}} - \tilde{\mathcal{H}}_{\bar{K}^{0}
\bar{K}^{0}} &\simeq &D \Biggl(-H^{w}_{K^{0} \bar{K}^{0}}
 \  \frac{d \Sigma_{\bar{K}^{0} K^{0}}(m)}{d m}\nonumber \\
 &&\;\;\;\;\;\;+H^{w}_{\bar{K}^{0} K^{0}}  \  \frac{d
\Sigma_{K^{0} \bar{K}^{0}}(m)}{d m}\Biggl)\neq 0 . \label{j2-72}
\end{eqnarray}

Note that from this last relation an important conclusion follows:
If  ${H}^{w}_{K^{0} \bar{K}^{0}} = 0, {H}^{w}_{\bar{K}^{0} K^{0}} =
({H}^{w}_{K^{0} \bar{K}^{0}})^{\ast} = 0$ then
$\tilde{\mathcal{H}}_{K^{0} K^{0}} -
\tilde{\mathcal{H}}_{\bar{K}^{0} \bar{K}^{0}}  \simeq 0$ to the very
good accuracy. So, if the first order  $|\Delta S| = 2$ transitions,
eg., $K^{0} \ \rightleftharpoons \ \bar{K}^{0}$, are forbidden for
interactions, $H^{w}$, responsible for the decays of neutral mesons,
then the difference of diagonal matrix elements of the more accurate
effective Hamiltonian than the LOY effective Hamiltonian,
$\mathcal{H}^{WW}$, equals zero. This means that CPT invariance test
based on the LOY theory relations (\ref{j2-26}) can  be no longer
considered as the CPT symmetry tests but rather as tests for an
existence of the interactions causing the first order $|\Delta S| =
2$ transitions (see also \cite{j2:Urbanowski-Acta-2004}).

The eigenvectors of $\tilde{\mathcal{H}}$ for the eigenvalues
\begin{equation}
\tilde{\mu}_{L(S)} = \tilde{m}_{L(S)} - \frac{i}{2}
\tilde{\Gamma}_{L(S)}, \label{j2-mu-LS}
\end{equation}
have the form \cite{j2:Khalfin,j2:7}
\begin{eqnarray}
|\tilde{K}_{L}\rangle=\tilde{\rho}_{L}\bigg(
|K^{0}\rangle-\alpha_{L}|\bar{K}^{0}\rangle\bigg)
\label{j2-74}
\end{eqnarray}
and
\begin{eqnarray}
|\tilde{K}_{S}\rangle=\tilde{\rho}_{S}\bigg(|K^{0}\rangle-\alpha_{S}
|\bar{K}^{0}\rangle\bigg),\label{j2-75}
\end{eqnarray}
where the parameters $\tilde{\rho}_{L}, \tilde{\rho}_{S}$ can be
chosen as the real parameters,
\begin{eqnarray}
\alpha_{L(S)}=\frac{\tilde{h}_{z}-(+) \
\tilde{h}}{\tilde{\mathcal{H}}_{K^{0} \bar{K}^{0}}},\label{j2-76}
\end{eqnarray}
and the definition of $\tilde{h}_{z}$ is given by (\ref{j2-57}), and
\begin{eqnarray}
\tilde{h}=\sqrt{(\tilde{h}_{z})^{2}+\tilde{\mathcal{H}}_{K^{0}
\bar{K}^{0}}\tilde{\mathcal{H}}_{\bar{K}^{0} K^{0}}}. \label{j2-78}
\end{eqnarray}

Sometimes one uses the following expression for vectors
$|\tilde{K}_{L}\rangle$ and $|\tilde{K}_{S}\rangle$,
\cite{j2:Gaillard,j2:Bilenkij}, \cite{j2:5}
--- \cite{j2:branco}, \cite{j2:7,j2:Urbanowski-2004},
\begin{equation} |\tilde{K}_{L(S)}\rangle \equiv
N_{L(S)}\, [ (1 + \epsilon_{l(s)})\,|K^{0} \rangle + ( -1)(1 -
{\epsilon}_{l(s)})\, |\bar{K}^{0} \rangle ] . \label{j2-Kl-Ks}
\end{equation}
This form of eigenvectors for the effective Hamiltonian is used in
many papers when possible departures from CP-- or CPT--symmetry in
the system considered are discussed. The following  parameters are
used to describe the scale of CP-- and possible  CPT  -- violation
effects \cite{j2:Gaillard,j2:Bilenkij}, \cite{j2:5}
--- \cite{j2:branco}, \cite{j2:7},
\begin{equation} \epsilon
\stackrel{\rm def}{=} \frac{1}{2} ( {\epsilon}_{s}  + {\epsilon}_{l}
) \;\;\;\;\;\;\;\;\;\delta \stackrel{\rm def}{=} \frac{1}{2} (
{\epsilon}_{s} - {\epsilon}_{l}  )  . \label{delta}
\end{equation}
Within the LOY theory of time evolution in the subspace of neutral
kaons, $\epsilon$ describes violations of CP--symmetry and $\delta$
is considered as a CPT--violating parameter.

It seems to be interesting to compare eigenvectors
$|\tilde{K}_{L}\rangle$, (\ref{j2-74}), and $|\tilde{K}_{S}\rangle$,
(\ref{j2-75}), for $\tilde{\mathcal{H}}$ with those corresponding to
them, (i.e, $|K_{S}\rangle$, (\ref{j2-32a}), and $|K_{L}\rangle$,
(\ref{j2-33a})), for $\mathcal{H}^{WW}$. To achieve this goal one
should rewrite suitable expressions (\ref{j2-76}), so as to get a
convenient form of $\alpha_{L(S)}$ allowing one to express it by
means of $r$, (\ref{j2-33b}). After some algebra one can rewrite
(\ref{j2-76}) as
\begin{eqnarray}
\alpha_{L(S)}=\tilde{g}-(+)
\tilde{r}\sqrt{1+\frac{\tilde{g}^{2}}{\tilde{r}^{2}}}. \label{j2-82}
\end{eqnarray}
where
\begin{eqnarray}
\tilde{r}=\sqrt{\frac{\tilde{\mathcal{H}}_{\bar{K}^{0}
K^{0}}}{\tilde{\mathcal{H}}_{K^{0} \bar{K}^{0}}}} \label{j2-80a}
\end{eqnarray}
and
\begin{eqnarray}
\tilde{g}=\frac{\tilde{h}_{z}}{\tilde{\mathcal{H}}_{K^{0}
\bar{K}^{0}}} \label{j2-80}
\end{eqnarray}

The conclusion that the more accurate approximation leads to a more
realistic description of properties of a physical system seems to be
obvious. For this reason it seems that expressing matrix elements of
the  more accurate $\tilde{\mathcal{H}}$ instead of
$\mathcal{H}^{WW}$ in terms of the parameters obtained from
experiments is justified. So, using the form (\ref{j2-Kl-Ks}) of
eigenvectors for $\tilde{\mathcal{H}}$,  matrix elements
$\tilde{\mathcal{H}}_{\alpha \beta}$ of $\tilde{\mathcal{H}}$ can be
expressed in terms of observables  $\epsilon_{L}$ and $\epsilon_{S}$
and $\tilde{\mu}_{L(S)}$ (see, eg. \cite{j2:dafne,j2:7}). By means
of this method one finds, eg. that
\begin{eqnarray}
\tilde{g} \,= \, \frac{\tilde{h}_{z}}{\tilde{\mathcal{H}}_{K^{0}
\bar{K}^{0}}} \equiv \frac{1}{2} \, \frac{\tilde{\mathcal{H}}_{K^{0}
K^{0}} - \tilde{\mathcal{H}}_{\bar{K}^{0} \bar{K}^{0}}}
{\tilde{\mathcal{H}}_{K^{0} \bar{K}^{0}}}
=\frac{2\delta}{(1+\epsilon_{l})(1+\epsilon_{s})} , \label{H_z:H_12}
\end{eqnarray}
and
\begin{eqnarray}
\tilde{r}=\sqrt{\frac{(1-\epsilon_{l})(1-\epsilon_{s})}{(1+\epsilon_{l})
(1+\epsilon_{s})}}.
\label{j2-82c}
\end{eqnarray}
Experimentally measured  values  of  parameters ${\epsilon}_{l},
{\epsilon}_{s}$  are  very  small  for neutral  kaons. So, assuming
\begin{equation}    |{\varepsilon}_{l}|     \ll     1,     \;     \;
|{\varepsilon}_{s}|  \ll  1,   \label{j2-e-ls<<1}
\end{equation}
one finds that
\begin{equation}
|\tilde{g} |\; \ll \; 1 \;\;\;\;{\rm and}\;\;\;\;\; |\tilde{r}| \;
\simeq\;1, \label{j2-g<<1}
\end{equation}
and thus
\begin{equation}
|\frac{\tilde{g}}{\tilde{r}}| \,\ll
 \, 1. \label{j2-g<<r}
\end{equation}
Therefore to a very good approximation
\begin{equation}
\sqrt{1+\frac{\tilde{g}^{2}}{\tilde{r}^{2}}} \;\simeq \; 1 \,+\,
\frac{1}{2}\;\frac{\tilde{g}^{2}}{\tilde{r}^{2}}.
\end{equation}
So, the relation (\ref{j2-82}) can be approximated by the following
one
\begin{eqnarray}
\alpha_{L(S)}\simeq \tilde{g}-(+) \  \tilde{r} \ \bigg(1+
\frac{\tilde{g}^{2}}{2 \ \tilde{r}^{2}}\bigg), \label{j2-84}
\end{eqnarray}
or, taking into account (\ref{j2-g<<r}), by
\begin{eqnarray}
\alpha_{L}\simeq -\tilde{r}+\tilde{g}, \;\;\;\;\;\;{\rm
and}\;\;\;\;\;\;\; \alpha_{S}\simeq \tilde{r}+\tilde{g}.
\label{j2-85}
\end{eqnarray}
Inserting (\ref{j2-85}) into formulae  (\ref{j2-74}) and
(\ref{j2-75}) results in the following expressions for
$|\tilde{K}_{L}\rangle, |\tilde{K}_{S}\rangle$,
\begin{eqnarray}
|\tilde{K}_{L}\rangle=\tilde{\rho}_{L} \bigg(|K^{0}\rangle+\tilde{r}
\ |\bar{K}^{0}\rangle\bigg)-\tilde{g} \  \tilde{\rho}_{L} \
|\bar{K}^{0}\rangle \label{j2-87}
\end{eqnarray}
and
\begin{eqnarray}
|\tilde{K}_{S}\rangle=\tilde{\rho}_{S}\bigg(|K^{0}\rangle-\tilde{r}
\ |\bar{K}^{0}\rangle\bigg)-\tilde{g} \ \tilde{\rho}_{S} \
|\bar{K}^{0}\rangle. \label{j2-88}
\end{eqnarray}
Next, let us note that
\begin{eqnarray}
(\tilde{r})^{2} = \frac{\tilde{\mathcal{H}}_{\bar{K}^{0}
K^{0}}}{\tilde{\mathcal{H}}_{K^{0} \bar{K}^{0}}} \equiv
\frac{\mathcal{H}^{WW}_{\bar{K}^{0} K^{0}}+R_{K^{0} K^{0}}(m)
\frac{d R_{\bar{K}^{0} K^{0}}(m)}{d m} + R_{\bar{K}^{0} K^{0}}(m)
\frac{d R_{\bar{K}^{0} \bar{K}^{0}}(m)}{d
m}}{\mathcal{H}^{WW}_{K^{0} \bar{K}^{0}}+R_{\bar{K}^{0}
\bar{K}^{0}}(m) \frac{d R_{K^{0} \bar{K}^{0}}(m)}{d m} + R_{K^{0}
\bar{K}^{0}}(m) \frac{d R_{K^{0} K^{0}}(m)}{d m}}, \label{j2-106}
\end{eqnarray}
(because $\mathcal{H}^{WW}_{K^{0} \bar{K}^{0}} \equiv R_{K^{0}
\bar{K}^{0}}(m)$ and $\mathcal{H}^{WW}_{\bar{K}^{0} K^{0}} \equiv
R_{\bar{K}^{0} K^{0}}(m)$ --- see (\ref{j2-H-WW-ab1}) ). This means
that neglecting terms of the type $R_{\alpha \beta}(m) \frac{d\,
R_{\alpha' \beta'}(m)}{d m}$, one finds
\begin{eqnarray}
\frac{\tilde{\mathcal{H}}_{\bar{K}^{0}
K^{0}}}{\tilde{\mathcal{H}}_{K^{0} \bar{K}^{0}}} \simeq
\frac{\mathcal{H}^{WW}_{\bar{K}^{0} K^{0}}}{\mathcal{H}^{WW}_{K^{0}
\bar{K}^{0}}}, \label{j2-107}
\end{eqnarray}
which enables one to draw the conclusion that the difference between
expressions (\ref{j2-80a}) and (\ref{j2-33b}) is almost negligibly
small, i.e, that
\begin{equation}
\tilde{r} \simeq r, \label{tilde-r=r} \end{equation}
to the very
good approximation. So, in our analysis expressions (\ref{j2-85})
can be replaced by the following one
\begin{eqnarray}
\alpha_{S(L)} \simeq  + (-) \  r+\tilde{g}. \label{j2-86a}
\end{eqnarray}
The conclusion following from this property is that formulae
(\ref{j2-87}) and (\ref{j2-88}) for eigenvectors
$|\tilde{K}_{L}\rangle, |\tilde{K}_{S}\rangle$ of $\mathcal{H}$ can
be rewritten using eigenvectors $|{K}_{L}\rangle, |{K}_{S}\rangle$,
(\ref{j2-32a}), (\ref{j2-33a}) of $\mathcal{H}^{WW}$ as follows
\begin{eqnarray}
|\tilde{K}_{L}\rangle= \frac{\tilde{\rho}_{L}}{ \rho^{WW}} \
|K_{L}\rangle-\tilde{g} \ \tilde{\rho}_{L} \ |\bar{K}^{0}\rangle
\label{j2-89}
\end{eqnarray}
and
\begin{eqnarray}
|\tilde{K}_{S}\rangle= \frac{\tilde{\rho}_{S}}{ \rho^{WW}} \
|K_{S}\rangle-\tilde{g} \ \tilde{\rho}_{S} \  |\bar{K}^{0}\rangle .
\label{j2-90}
\end{eqnarray}

To complete our analysis we should remove $|\bar{K}^{0}\rangle$ from
(\ref{j2-89}) and (\ref{j2-90}). Expressing $|\bar{K}^{0}\rangle$ by
$|K_{S}\rangle$, (\ref{j2-32a}), and $|K_{L}\rangle$,
(\ref{j2-33a}), and then inserting such obtained formula for
$|\bar{K}^{0}\rangle$ into (\ref{j2-89}) and (\ref{j2-90}) yields
\begin{eqnarray}
|\tilde{K}_{S}\rangle=\frac{\tilde{\rho}_{S}}{\rho^{WW}}
\bigg[\bigg(1+\frac{\tilde{g}}{2 r}\bigg)|K_{S}\rangle-
\frac{\tilde{g}}{2 r}|K_{L}\rangle\bigg] \label{j2-101}
\end{eqnarray}
and
\begin{eqnarray}
|\tilde{K}_{L}\rangle=\frac{\tilde{\rho}_{L}}{\rho^{WW}}
\bigg[\bigg(1-\frac{\tilde{g}}{2 r}\bigg)|K_{L}\rangle+
\frac{\tilde{g}}{2 r}|K_{S}\rangle\bigg]. \label{j2-102}
\end{eqnarray}

Similarly, starting from relations (\ref{j2-87}) and (\ref{j2-88})
one can express $|K^{0}\rangle$ and $|\bar{K}^{0}\rangle$ by means
of eigenvectors $|\tilde{K}_{L}\rangle, |\tilde{K}_{S}\rangle$ for
$\tilde{\mathcal{H}}$. Next inserting  expressions obtained in this
way for $|K^{0}\rangle, |\bar{K}^{0}\rangle$ into formulae for
$|K_{S}\rangle$, (\ref{j2-32a}), and $|K_{L}\rangle$,
(\ref{j2-33a}), and using property (\ref{tilde-r=r}) one finds that
\begin{eqnarray}
|K_{S}\rangle\,\simeq\,\rho^{WW}\,
\bigg[\frac{1}{\tilde{\rho}_{S}}\, \bigg(1-\frac{\tilde{g}}{2
r}\bigg)|\tilde{K}_{S}\rangle \,+\, \frac{1}{\tilde{\rho}_{L}}\,
\frac{\tilde{g}}{2 r}|\tilde{K}_{L}\rangle \bigg], \label{j2-101-vv}
\end{eqnarray}
and
\begin{eqnarray}
|K_{L}\rangle\,\simeq\,\rho^{WW}\,\bigg[\frac{1}{\tilde{\rho}_{L}}\,
\bigg(1\,+\,\frac{\tilde{g}}{2 r}\bigg)|\tilde{K}_{L}\rangle\,-\,
\frac{1}{\tilde{\rho}_{S}}\, \frac{\tilde{g}}{2
r}|\tilde{K}_{S}\rangle \bigg]. \label{j2-102-vv}
\end{eqnarray}

All the above considerations are carried out within the assumption
that the system containing neutral kaons is CPT invariant. This
means that contrary to the one of fundamental results of the LOY
theory (see (\ref{j2-KS-KL})) the scalar product of vectors
eigenvectors $|\tilde{K}_{S}\rangle $ and $|\tilde{K}_{L}\rangle$
for the Khalfin's improved effective Hamiltonian
$\tilde{\mathcal{H}}$ can not be real in the CPT invariant system.
Indeed from (\ref{j2-101}) and (\ref{j2-102}) it follows that
\begin{eqnarray}
\langle\tilde{K}_{S}|\tilde{K}_{L}\rangle&=&\nonumber
\frac{\tilde{\rho}_{S}\tilde{\rho}_{L}}{(\rho^{WW})^{2}}\,
\bigg[\bigg(1\,-\, 2i\, \Im\,(\frac{\tilde{g}}{2
r})\,-\,\frac{|\tilde{g}|^{2}}{2|r|^{2}}\bigg)\langle
K_{S}|K_{L}\rangle
\nonumber \\
&&+\,2i \, \Im\,(\frac{\tilde{g}}{2 r}) \,+\,
\frac{|\tilde{g}|^{2}}{2|r|^{2}} \bigg]   \label{j2-103}\\
&\neq &(\langle\tilde{K}_{S}|\tilde{K}_{L}\rangle)^{\ast} =
\langle\tilde{K}_{L}|\tilde{K}_{S}\rangle , \nonumber
\end{eqnarray}
where $\Im\,(z)$ denotes the imaginary part of the complex number
$z$. (Note that according to our assumptions, parameters
$\tilde{\rho}_{S},$ $\tilde{\rho}_{L}$ and $\rho^{WW}$ are the real
numbers and the product $\langle K_{S}|K_{L}\rangle$ is  real too
--- see (\ref{j2-KS-KL})).

Ignoring in (\ref{j2-103}) terms of order
$\frac{|\tilde{g}|^{2}}{|r|^{2}}$ gives
\begin{eqnarray}
\langle\tilde{K}_{S}|\tilde{K}_{L}\rangle \simeq
\frac{\tilde{\rho}_{S}\tilde{\rho}_{L}}{(\rho^{WW})^{2}}
\bigg[\bigg(1\,-\,2i \, \Im\,(\frac{\tilde{g}}{2 r}) \bigg)\langle
K_{S}|K_{L}\rangle\, +\,2i \, \Im\,(\frac{\tilde{g}}{2 r})
 \bigg], \label{j2-104}
\end{eqnarray}
to an accuracy sufficient for our analysis. Note that these last two
relations are in perfect agreement with the result obtained in
\cite{j2:Urbanowski-2004}, where the general proof that the scalar
product of eigenvectors for the exact effective Hamiltonian
corresponding to short and long living superpositions of $K^{0}$ and
$\bar{K}^{0}$ mesons can not be real in CPT invariant system is
given.

\section{Final remarks}

At the beginning of this Section we should explain why the final
formulae for matrix elements of the improved effective Hamiltonian
$\tilde{\mathcal{H}}$ derived in Sec. 3 differ a little from those
obtained by Khalfin in his paper \cite{j2:Khalfin}. This is because
in \cite{j2:Khalfin} some additional assumptions were used to
simplify the calculations. A detailed analysis of these assumptions
and recent experimental data suggest that some of them can not be
considered as universally valid for neutral meson complex. Deriving
our formulae, ie. (\ref{j2-48}), (\ref{j2-51}) (and (\ref{j2-48}),
(\ref{j2-51})), for $\tilde{\mathcal{H}}_{\alpha \beta}$,  only the
general Khalfin's assumptions having the form (\ref{j2-40}),
(\ref{j2-42}) have been used. The general form of
$\tilde{\mathcal{H}}$ obtained in Sec. 3 and given by formula
(\ref{j2-47}) and the formula for $\tilde{\mathcal{H}}$ derived in
\cite{j2:Khalfin} are identical. On the other hand our formulae
(\ref{j2-48}), (\ref{j2-51}) for $\tilde{\mathcal{H}}_{\alpha
\beta}$ seem to be more general  and more accurate than those
obtained in \cite{j2:Khalfin}.

Note that if CP symmetry is conserved then $H^{w}_{K^{0}
\bar{K}^{0}} = H^{w}_{\bar{K}^{0} K^{0}}$ and relations
(\ref{j2-53}) are also valid. Then from (\ref{j2-57}) it follows
that $(\tilde{\mathcal{H}}_{K^{0} K^{0}} -
\tilde{\mathcal{H}}_{\bar{K}^{0} \bar{K}^{0}}) = 0$ in CP invariant
system. This means that in such a case a picture of neutral meson
system obtained within the use of the Khalfin's improved effective
Hamiltonian, $\tilde{\mathcal{H}}$, is the same as that obtained
using the LOY effective hamiltonian $\tilde{\mathcal{H}}^{WW}$.

Let us now focus our attention on the case of violated CP and
conserved CPT symmetries. The most important conclusion following
from the Sec. 2 --- Sec. 4 is that the minimal improvement of the WW
approximation leads to such properties of the new effective
Hamiltonian for the neutral kaon complex which contradict standard
predictions of the LOY theory for systems in which CP symmetry is
violated. Relations (\ref{j2-56}), (\ref{j2-57}), (\ref{j2-72}) are
an example of such properties. These relations state that the
standard result of the LOY theory that diagonal matrix elements of
the effective Hamiltonian governing the time evolution of neutral
kaons are equal if the system containing these neutral $K$ mesons is
CPT invariant, can not be considered as the true property for  the
real systems. Another example is the relation (\ref{j2-103}). This
result obtained for the eigenvectors of the Khalfin's improved
effective Hamiltonian shakes another standard prediction of the LOY
theory, i.e., the property (\ref{j2-KS-KL}), that the scalar product
of eigenvectors for the effective Hamiltonian should be a real
number for system preserving CPT symmetry. All these corrections to
the corresponding LOY results are very small. They are of order of
the parameter $\tilde{g}$ defined by relations (\ref{j2-80}) and
(\ref{j2-57}). Nevertheless they all have a nonzero value. This
means that the LOY theory interpretation of experimentally measured
parameters for neutral meson complexes may not reflect properly all
real properties of such complexes. For example, properties
(\ref{j2-56}), (\ref{j2-57}), (\ref{j2-72}) mean by (\ref{delta})
and (\ref{H_z:H_12}) that there must be $ \epsilon_{l} \neq
\epsilon_{s}$ when CPT symmetry holds and CP is violated (see also
\cite{novikov}). Of course this conclusion contradicts the standard
predictions of the LOY theory.

One more observation is in agreement with last conclusions. Namely
assuming that picture of the physical system given by parameters
calculated within the more accurate approximation is more realistic
than that followed from the less accurate one, from
(\ref{j2-101-vv}), (\ref{j2-102-vv}) we can draw the following
conclusion: Superpositions of states $|K^{0}\rangle,
|\bar{K}^{0}\rangle$ of type $|K_{L}\rangle, |K_{S}\rangle$, having
the form (\ref{j2-32a}), (\ref{j2-33a}), with expansion coefficients
$r$, (\ref{j2-33b}), calculated within the WW approximation can not
be considered as the real physical states. Relations
(\ref{j2-101-vv}), (\ref{j2-102-vv}) show that vectors
$|K_{L}\rangle$, (\ref{j2-32a}), and  $|K_{S}\rangle$,
(\ref{j2-33a}), are linear combinations of eigenvectors
$|\tilde{K}_{L}\rangle, |\tilde{K}_{S}\rangle$ for the more accurate
$\tilde{\mathcal{H}}$ than $\tilde{\mathcal{H}}^{WW}$. It seems that
rather vectors $|\tilde{K}_{L}\rangle, |\tilde{K}_{S}\rangle$ as the
eigenvectors of the more accurate effective Hamiltonian should
pretend to representing physical states of the neutral kaons.

It seems to be interesting that the relation (\ref{j2-72}) confirms
the result obtained in \cite{j2:Urbanowski-Acta-2004} within the use
of a different formalism than that leading to the result
(\ref{j2-72}) in Sec. 3. This result and the result discussed in
\cite{j2:Urbanowski-Acta-2004} suggest that tests for neutral meson
complexes  based on the measurement of the difference of the
diagonal matrix elements of the effective Hamiltonian for such
complex can not be considered as the CPT invariance tests. From
(\ref{j2-72}) it follows that this difference equals zero in CPT
invariant system  if $\langle K^{0}|H^{w}|\bar{K}^{0}\rangle = 0$
and does not equal zero if $\langle K^{0}|H^{w}|\bar{K}^{0}\rangle
\neq 0$. This means that this difference can not be  equal to zero
if the first order $|\Delta S| = 2$ transitions $K^{0}
\rightleftharpoons \bar{K}^{0}$ take place in the system considered.
So the tests mentioned should rather be considered as the tests for
existence of the interactions  causing the first order $|\Delta S| =
2$ transitions $K^{0} \rightleftharpoons \bar{K}^{0}$ in  the
system.

Note also that the properties (\ref{j2-57}) and (\ref{j2-103}) of
the Khalfin's effective Hamiltonian are in perfect agreement with
the analogous rigorous results obtained in \cite{j2:plb-2002} and
\cite{j2:Urbanowski-2004} without the use of any approximations for
the exact effective Hamiltonian.

It is also interesting to compare solutions (\ref{j2-23}) of the LOY
evolution equation  (\ref{j2-25}) for the subspace
$\mathfrak{H}_{\parallel}$ with solutions (\ref{j2-44}) of the
evolution equation (\ref{LOY-improved-eq}) containing Khalfin's
improved effective Hamiltonian $\tilde{\mathcal{H}}$. We already
have a solution  $|K^{0}_{WW}(t)\rangle_{\parallel}$, (\ref{j2-36}),
of  Eq. (\ref{j2-25}). Let us find an analogous solution
$|\tilde{K}^{0}(t)\rangle_{\parallel}$ of Eq.
(\ref{LOY-improved-eq}). From (\ref{j2-44}) one finds
\begin{equation}
|K^{0}(t)\rangle_{\parallel} \, \simeq \, |\tilde{K}^{0}(t)
\rangle_{\parallel} = e^{\textstyle -i \tilde{\mathcal{H}} t}
  \  |\tilde{K^{0}}(0) \rangle_{\parallel}, \;\;\;\;(t >0), \label{K(t)-imp}
\end{equation}
where according to (\ref{a-0-tilde}),
\begin{equation}
|\tilde{K^{0}}(0) \rangle_{\parallel} = \mathbf{A} \ |K^{0} \rangle
\equiv a_{11}\, |K^{0}\rangle + a_{21}\, |\bar{K}^{0}\rangle,
\label{K(0)-tilde}
\end{equation}
and $a_{jk}, (j,k =1,2)$ are matrix elements of $\mathbf{A}$ (see
(\ref{a-0-tilde}), (\ref{j2-45})). Thus
\begin{equation}
e^{\textstyle -i \tilde{\mathcal{H}} t}
  \  |\tilde{K^{0}}(0) \rangle_{\parallel} \equiv
a_{11} \; e^{\textstyle -i \tilde{\mathcal{H}} t}
  \  |K^{0}\rangle + a_{21} \;e^{\textstyle -i \tilde{\mathcal{H}} t}
  \  |\bar{K}^{0} \rangle. \label{K(t)-imp-1}
\end{equation}
Next using relations (\ref{j2-74}), (\ref{j2-75}) one can express
vectors $|K^{0} \rangle, |\bar{K}^{0}\rangle$ by means of
eigenvectors $|\tilde{K}_{L}\rangle$ and $|\tilde{K}_{S}\rangle$ for
$\tilde{\mathcal{H}}$. The result of the action of $e^{\textstyle -i
\tilde{\mathcal{H}} t}$ onto $|\tilde{K}_{L}\rangle,
|\tilde{K}_{S}\rangle$ can be easily found. Having this result one
can return to the base vectors $|K^{0} \rangle, |\bar{K}^{0}\rangle$
which yields
\begin{eqnarray}
|\tilde{K}^{0}(t)\rangle_{\parallel} & = & \;\;\;
\frac{1}{\alpha_{S} - \alpha_{L}} \; \Big[ (a_{11}\,\alpha_{S}\,
+\,a_{21})\, e^{\textstyle -i(\tilde{m}_{L} \
- \ \frac{i}{2} \ \tilde{\Gamma}_{L})t}  \nonumber\\
&& \makebox[20pt]{} - (a_{11}\,\alpha_{L}\, +\,a_{21})\,
e^{\textstyle -i(\tilde{m}_{S} \ - \ \frac{i}{2} \
\tilde{\Gamma}_{S})t}\Big] |K^{0} \rangle  \nonumber\\
&& -\frac{1}{\alpha_{S} - \alpha_{L}} \; \Big[ (a_{11}\,\alpha_{S}\,
+\,a_{21})\;\alpha_{L}\; e^{\textstyle -i(\tilde{m}_{L} \
- \ \frac{i}{2} \ \tilde{\Gamma}_{L})t} \nonumber\\
&& \makebox[20pt]{} - (a_{11}\,\alpha_{L}\, +\,a_{21})\;
\alpha_{S}\; e^{\textstyle -i(\tilde{m}_{S} \ - \ \frac{i}{2} \
\tilde{\Gamma}_{S})t}\Big] |\bar{K}^{0} \rangle.
\label{K(t)-imp-final}
\end{eqnarray}

Note that this last expression is not equal to the  analogous
formula obtained in \cite{j2:Khalfin}. The cause of this is
described in Sec. 4 after formulae (\ref{j2-48}), (\ref{j2-51}).
Nevertheless the general conclusions following from
(\ref{K(t)-imp-final}) and from the mentioned corresponding
Khalfin's formula are similar. Namely from (\ref{j2-36}) and from
(\ref{K(t)-imp-final}) (as well as from the analogous formula
obtained in \cite{j2:Khalfin}) it follows that
\begin{eqnarray*}
|\langle K^{0}|K^{0}_{WW}(t)\rangle_{\parallel}|^{2} &\neq& |\langle
K^{0}|\tilde{K}^{0}(t)\rangle_{\parallel}|^{2} , \label{prob-1} \\
|\langle \bar{K}^{0}|K^{0}_{WW}(t)\rangle_{\parallel}|^{2} &\neq&
|\langle \bar{K}^{0}|\tilde{K}^{0}(t)\rangle_{\parallel}|^{2} .
\label{prob-2}
\end{eqnarray*}

Thus within the Khalfin's improved theory of neutral mesons formulae
describing the strangeness (particle--antiparticle) oscillations
lead to some corrections to the corresponding standard predictions
of the LOY theory. This effect was called in \cite{j2:Khalfin} "a
new CP violation effect". So if more accurate tests of the mentioned
oscillations detect some departures from the predictions obtained
within the use probabilities  $|\langle
K^{0}|K^{0}_{WW}(t)\rangle_{\parallel}|^{2}$ and $|\langle
\bar{K}^{0}|K^{0}_{WW}(t)\rangle_{\parallel}|^{2}$ then such an
effect should be considered as a very probable confirmation of the
improved Khalfin's theory of neutral mesons.

The complementary conclusion to the above one is the observation
following from the results described in Sec. 4: More accurate
measurements of the difference $(\tilde{\mathcal{H}}_{K^{0} K^{0}} -
\tilde{\mathcal{H}}_{\bar{K}^{0} \bar{K}^{0}})$, that is of the
parameters $\epsilon_{L}, \epsilon_{S}$ and $\delta$, (\ref{delta}),
(see (\ref{H_z:H_12}) and (\ref{j2-57})) should also make possible
to see the difference between the standard predictions of the LOY
theory and the theory based on the more accurate effective
Hamiltonian derived by Khalfin.

The investigations of the above mentioned "a new CP violation
effect" and related problems by means of the another and more
general method than that used in \cite{j2:Khalfin} were continued in
\cite{Khalfin-1,Khalfin-2,Khalfin-3}. The expected form of the
probabilities $|\langle
K^{0}|\tilde{K}^{0}(t)\rangle_{\parallel}|^{2}$ and $|\langle
\bar{K}^{0}|\tilde{K}^{0}(t)\rangle_{\parallel}|^{2}$ is given in
\cite{Khalfin-2,Khalfin-3}.

The last comment. In \cite{j2:ijmpa-1993} (see also \cite{j2:7}) a
new approach to describing time evolution in neutral kaon complex
was proposed. This approach is based on the exact evolution equation
for subspace of states, $\mathfrak{H}_{||}$ describing neutral
mesons sometimes called the Kr\'{o}likowski--Rzeuwski (KR) equation
for the distinguished component of a state vector (see
\cite{j2:KR-1,j2:KR-2}). Using KR equation the approximate effective
Hamiltonian for $H_{||}$ neutral meson complex was derived. The
approximation used there has the advantage on the WW and LOY
approximations that all steps leading to the final formulae for the
approximate $H_{||}$ are well defined. Such obtained $H_{||}$
differs from $\mathcal{H}^{WW}$ and its matrix elements have the
form closed to that of $\tilde{\mathcal{H}}$. In details: The
property of $\tilde{\mathcal{H}}$ of type (\ref{j2-57}) occurs also
for the difference of the diagonal matrix elements of the $H_{||}$.
Replacing $D$ in the formula (\ref{j2-72}) by $D \simeq \mathbb{I}$
(which for some purposes is an sufficient approximation ) the
relation (\ref{j2-72}) becomes identical with the analogous relation
derived in \cite{j2:Urbanowski-Acta-2004} for the $H_{||}$. The
property (\ref{j2-103}) of the scalar product of the eigenvectors
for $\tilde{\mathcal{H}}$ takes place also for the eigenvectors of
the approximate $H_{||}$ obtained within the use of KR equation.

\end{document}